\documentclass[epj,nopacs]{svjour}

\usepackage{comment}
\usepackage{graphicx}
\usepackage{subfigure} 
\usepackage{microtype}
\usepackage{hyperref}

\begin{document}

\date{\today}

\title{A high-pressure hydrogen time projection chamber for the MuCap experiment}

\author{
{J.\,Egger}\inst{1} \and
{D.\,Fahrni}\inst{1} \and
{M.\,Hildebrandt}\inst{1} \and
{A.\,Hofer}\inst{1} \and
{L.\,Meier}\inst{1} \and
{C.\,Petitjean}\inst{1} \and
{V.A.\,Andreev}\inst{2} \and
{T.I.\,Banks}\inst{3} \and
{S.M.\,Clayton}\inst{4} \and
{V.A.\,Ganzha}\inst{2} \and
{F.E.\,Gray}\inst{5,6,3} \and
{P.\,Kammel}\inst{6,4} \and
{B.\,Kiburg}\inst{6,4} \and
{P.A.\,Kravtsov}\inst{2} \and
{A.G.\,Krivshich}\inst{2} \and
{B.\,Lauss}\inst{1,3} \and
{E.M.\,Maev}\inst{2} \and
{O.E.\,Maev}\inst{2} \and
{G.\,Petrov}\inst{2} \and
{G.G.\,Semenchuk}\inst{2} \and
{A.A.\,Vasilyev}\inst{2} \and
{A.A.\,Vorobyov}\inst{2} \and
{M.E.\,Vznuzdaev}\inst{2} \and
{P.\,Winter}\inst{6,4} 
}

\institute{
Paul Scherrer Institute, CH-5232 Villigen PSI, Switzerland \and
Petersburg Nuclear Physics Institute, Gatchina 188350, Russia \and
Department of Physics, University of California, Berkeley, California 94720, USA \and
Department of Physics, University of Illinois at Urbana-Champaign, Urbana, Illinois 61801, USA \and
Department of Physics and Computational Science, Regis University, Denver, Colorado 80221, USA \and
Department of Physics, University of Washington, Seattle, Washington  98195, USA
}

\abstract{
The MuCap experiment at the Paul Scherrer Institute performed a high-precision 
measurement of the rate of the basic electroweak process of nuclear muon 
capture by the proton, $\mu^- + p \rightarrow n + \nu_\mu$.
The experimental approach was based on the use of a time projection 
chamber~(TPC) that operated in pure hydrogen gas at a pressure of 10~bar 
and functioned as an active muon stopping target.  The TPC detected the tracks 
of individual muon arrivals in three dimensions, while the trajectories of 
outgoing decay~(Michel) electrons were measured by two surrounding wire 
chambers and a plastic scintillation hodoscope. The muon and electron 
detectors together enabled a precise measurement of the $\mu p$ atom's 
lifetime, from which the nuclear muon capture rate was deduced.  
The TPC was also used to monitor the purity of the hydrogen gas by detecting the nuclear recoils that follow muon capture by elemental impurities.
This paper describes the TPC design and performance in detail.
}


\maketitle


\section{Introduction}
\label{sec:intro}

The MuCap experiment~\cite{MuCapTP:1996,MuCapTP:2001,Andreev:2007wg,Andreev:2012fj} 
measured the rate $\Lambda^{}_S = 714.9 \, \pm \, 5.4_{\rm stat} \, \pm \, 5.1_{\rm syst}$~s$^{-1}$ 
for the basic electroweak process of nuclear muon capture by the proton,
$\mu^- + p \to n + \nu_\mu$.  This unprecedented precision determined 
the least well known of the nucleon's charged-current form factors, the induced
pseudoscalar coupling \mbox{${\textsl g}^{}_P$}, to 6.8\%~\cite{Bernard:1994wn,Bernard:2001rs,Czarnecki:2007th}.  MuCap 
collected data from 2004--2007 at the Paul Scherrer Institute~(PSI) in 
Switzerland using a time projection chamber~(TPC; see~\cite{Hilke:2010} 
for a review) to detect low-energy muons ($p\approx$~35~MeV/$c$) arriving 
from the facility's $\pi$E3 beamline.  Tests of early prototype chambers were 
reported in 2001~\cite{Maev:2001,Maev:2001yx} and 2002~\cite{Maev:2002sn}. 
The experiment's first result was published in 2007~\cite{Andreev:2007wg}, 
based on data collected in 2004.
An abbreviated paper on the final TPC design was published in 
conference proceedings in 2011~\cite{Egger:2011zz}.
The experiment's final results, based on data collected in 2006 and 2007, 
were published in 2013~\cite{Andreev:2012fj}.  

\subsection{Experimental requirements}
\label{ssec:exp-requirements}

%
The most precise method for determining the capture rate \mbox{$\Lambda^{}_S$} 
from the singlet ground state of the $\mu p$ atom is the lifetime technique.
In this method, the $\mu p$ bound-state disappearance rate $\lambda^-$ 
and the free $\mu^+$ decay rate $\lambda^+$ are determined separately by 
measuring the time spectrum of decay electrons from each.  The singlet
capture rate is deduced from the 
difference \mbox{$\Lambda^{}_S = \lambda^- - \lambda^+$}.  Other methods, such 
as direct measurement of the absolute neutron yield from nuclear captures in 
the $\mu p$ system, are not able to achieve a comparable precision.  
However, in order to make a precision measurement using the lifetime technique,
the experimental setup had to be capable of collecting very high statistics---namely,  $10^{10}$ muon decay events for a measurement of 
$\lambda^-$ to 10 parts per million (ppm).  When combined with the 1 ppm
measurement of $\lambda^+$ from the MuLan experiment~\cite{Tishchenko:2012ie}, 
this level of statistics determined \mbox{$\Lambda^{}_S$} to 1\%.

 The measurement had to be performed in low-density hydrogen gas 
in order to minimize distortions to the $\mu p$ lifetime from formation 
of $p\mu p$ muonic molecules~\cite{Andreev:2007wg}.  The MuCap target gas 
density was $\sim$~1\% of liquid hydrogen; this corresponds to a gas pressure 
of $\sim$~10 bar at 300~K. 

 The hydrogen gas had to be kept very clean, with concentrations of other 
elements $\leq 10$~ppb.  This stringent requirement was necessary because a 
muon will preferentially transfer from the $\mu p$ atom to a heavier 
element~$Z$ due to its larger binding energy, and the capture rates 
for $\mu p$ and $\mu Z$ systems can differ by 
orders of magnitude due to the fact that the nuclear muon capture rate 
increases with $\sim Z^4$.  
Using ultra-pure hydrogen gas therefore reduced unwanted contributions 
to the $\mu^-$ lifetime spectrum from nuclear captures by impurities.
The TPC itself was used to monitor the purity of the gas, using a technique
described in sect.~\ref{sssec:impurities}.

It was essential to clearly identify muons that stopped in the hydrogen 
gas target because accepting muons that stopped in $Z>1$ materials 
would have distorted the lifetime spectrum for the reasons described above.

The hydrogen gas had to be depleted of deuterium, a naturally abundant 
isotope (150~ppm), to $<$~100~ppb.  Such isotopically pure hydrogen is 
commonly referred to as protium.   The presence of deuterium introduces a potential systematic error because a muon
will preferentially transfer from the $\mu p$ atom to the more tightly bound
deuterium atom, and the resulting $\mu d$ atom can diffuse large distances due 
to a Ramsauer-Townsend 
minimum in the $\mu d + p$ cross section.  As a result, the muon would 
be lost to the data analysis in a time-dependent manner,
or possibly transported to surrounding detector materials to
undergo capture there.  Either of these possible effects would distort the 
measured lifetime.


\subsection{Detector concept}
\label{ssec:tpc-concept}

To simultaneously satisfy the experiment's many technical requirements, an 
apparatus (fig.~\ref{fig:exp-setup}) containing three main functional groups 
was constructed:
\begin{figure}[tp]
  \begin{center}
  \includegraphics[width=\columnwidth]{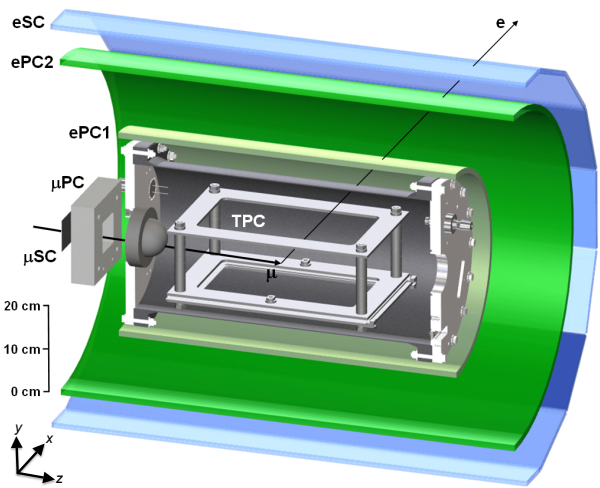}
  \caption{Drawing of the MuCap apparatus in cross section.  Beamline muons arrived
  from the left, passed through a series of entrance 
  detectors ($\mu$SC, $\mu$PC), and stopped in the TPC inside the 
  hydrogen-gas-filled pressure vessel.  Decay electrons were detected by the 
  surrounding wire chambers and hodoscope.  The pressure vessel's downstream 
  support structure and gas system are not shown.  (A version of this figure
  was previously published in~\cite{Andreev:2007wg}.)}
  \label{fig:exp-setup}
  \end{center}
\end{figure}
\begin{enumerate}
\item A {muon beam telescope}, which registered the arrival time of each muon and 
thus the start of its lifetime clock.  The telescope consisted of a plastic scintillation
counter, $\mu$SC, for fast timing
and an $x$-$y$ grid wire chamber, $\mu$PC, to provide position information
and a measure of detection redundancy.
Preceding this telescope was a veto scintillator, $\mu$SCA (not shown), for beam collimation; it had a 35-mm-diameter
hole and was followed by a sheet of lead of the same shape to ensure that errant muons were captured quickly.
\item The {TPC}, which operated in pure protium gas at 10~bar inside 
a cylindrical pressure vessel and tracked incoming muon trajectories 
in three dimensions.
\item A series of {concentric electron detectors}, which surrounded the pressure 
vessel and measured the emission time and trajectory of outgoing decay electrons, 
with $\sim$~$3 \pi$ solid-angle coverage.  
The electron detectors consisted of two nested wire chambers, ePC1 and ePC2, and 
an outer plastic scintillation hodoscope, eSC.  The ePC detectors provided information 
about the three-dimensional trajectory of the decay electron, while the eSC provided
fast timing information for the stop of the muon lifetime clock.
\end{enumerate}
A description of the MuCap physics results can be found in 
refs.~\cite{Andreev:2007wg,Andreev:2012fj}, and the experiment is
described in detail in refs.~\cite{Banks:PhD,Clayton:PhD,Kiburg:PhD,Knaack:PhD}.
This paper focuses primarily on describing the TPC and related infrastructure. 

Located in the center of the apparatus, the TPC was the crucial detector that 
made a high-precision muon capture rate measurement possible.  Its primary 
purpose was to detect the trajectory of each incoming muon and thereby 
enable the unambiguous identification of muons that stopped in the protium 
gas.  In this way, unwanted contributions to the decay time spectrum 
from muon stops in surrounding $Z>1$ materials were dramatically 
suppressed.  The TPC and ePCs in conjunction enabled reconstruction 
of each $\mu$-$e$ decay event vertex, providing a powerful method for 
suppressing accidental backgrounds.  The TPC was a versatile instrument 
which could also register signals from exotic processes such as nuclear 
muon capture.  Indeed, its response to
the nuclear recoils from captures by $Z>1$ elements in the protium gas 
proved essential during data taking as a means for in-situ monitoring of 
the contamination levels.

\subsection{Hydrogen wire chambers}
\label{sec:hydrogen}

The first TPC (detector IKAR) operating in pure hydrogen was
developed in Gatchina~\cite{Vorobyov1974509}, and 
it was successfully used in the Coulomb interference experiments WA9 and NA8  
at CERN~\cite{Burq1983285}.  This TPC operated in ionization mode  
without gas amplification.  The same mode was used later in the
MAYA~\cite{Demonchy2007341} and $\mu$CF~\cite{muCF} experiments.
The MuCap TPC was designed to operate with gas amplification in 
pure hydrogen, a regime in which there was little world experience.
TREAD, a cylindrical TPC operating in hydrogen
at a pressure of 15~atm, was used to study photon diffraction dissociation 
in Fermilab E612~\cite{Chapin:1981zz,Chapin:1984uk}; it was the first
TPC used with pure hydrogen that had gas gain.  We are not aware of any
others, except for a TPC that was demonstrated as a directional neutron 
detector, designed to monitor fissile material for nuclear security 
applications~\cite{Jovanovic}.

Pure hydrogen is a difficult chamber 
gas with no quenching of ultraviolet photons.  High gas pressures 
require higher voltages at the amplification stage than 
conventional chambers.  The TREAD group parameterized and confirmed 
measurements~\cite{PhysRev.59.850,PhysRev.104.273} of the first Townsend ionization 
coefficient $\alpha$, which characterizes the electron multiplication in the 
Townsend avalanche ($dn/n = \alpha \, dx$).   They fit the data to the 
formula~\cite{Chapin:1984uk}
\begin{equation} 
\label{eq:townsend}
\alpha / p = A \cdot e^{-b p / E} ~,
\end{equation}
where $p$ is the pressure, $E$ is the electric field magnitude, 
$A = 5.1$~cm$^{-1}$~Torr$^{-1}$ and $b = 138.8 \pm 0.4$~V~cm$^{-1}$~Torr$^{-1}$.
This parameterization agrees well with the values calculated by the 
Magboltz~\cite{Biagi:1999nwa} program within the
Garfield~\cite{Veenhof} tool that we used to simulate the MuCap TPC.
At a pressure of 10~bar as in MuCap, $\alpha$ rises steeply 
through 1~cm$^{-1}$ at $E \sim 10^{5}$~V/cm; this is therefore the critical 
field to trigger gas amplification.

\section{Apparatus}
\label{sec:apparatus}

\subsection{Design}
\label{ssec:design}

A prototype MuCap TPC of near full size was built in Gatchina, Russia, in 1997 
and tested from 1998--2000 in a muon beamline at 
PSI~\cite{Maev:2001,Maev:2001yx,Maev:2002sn}. 
These tests demonstrated the feasibility of 
stopping a muon beam inside a TPC operating in 10~bar hydrogen.
While the experimental setup did not meet the gas purity requirements for MuCap, 
the prototype TPC provided information essential to the design and construction 
of the next detector.  

\begin{figure}[tp]
  \begin{center}
  \includegraphics[width=\columnwidth]{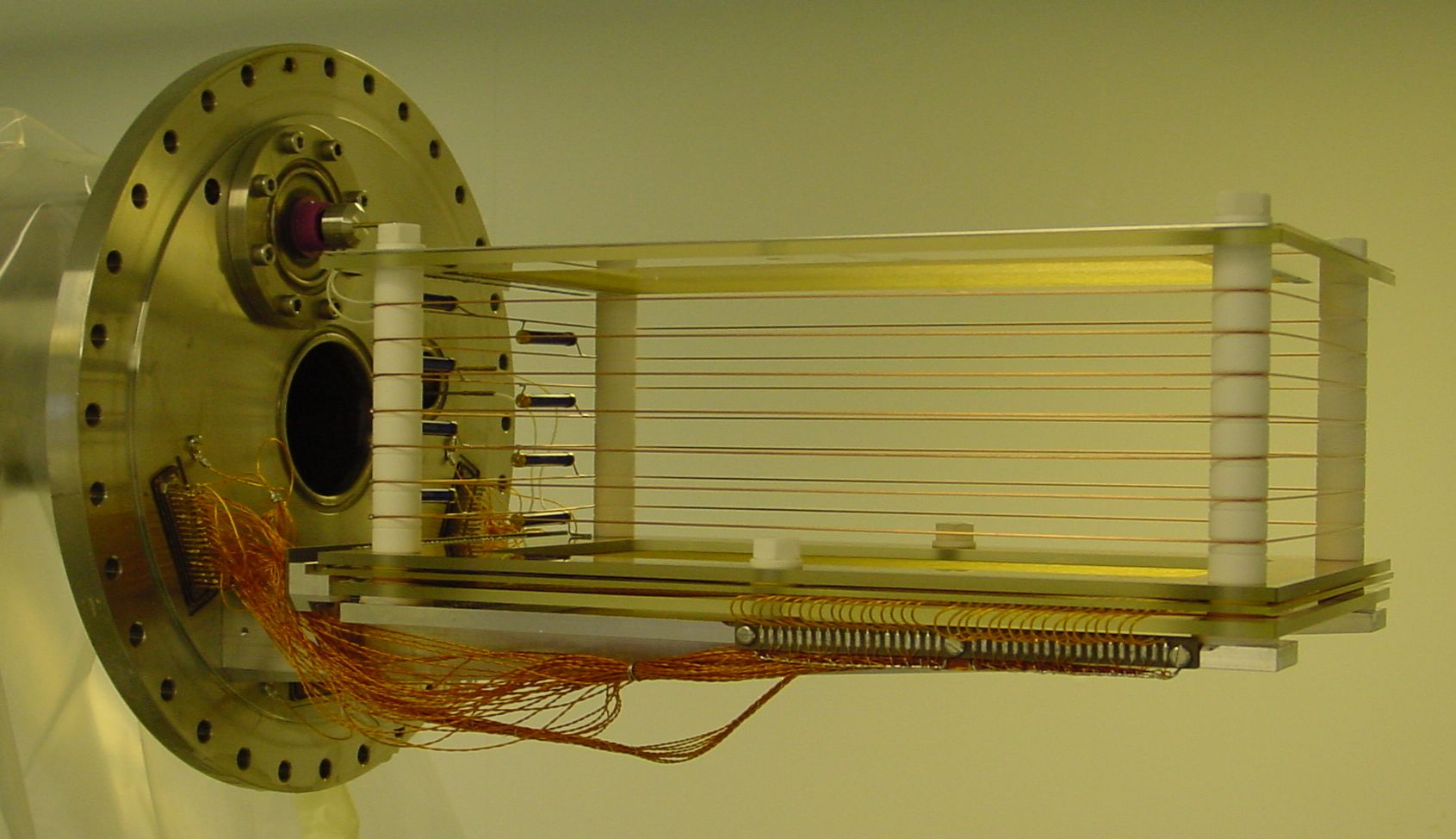}
  \caption{Side view of the TPC after its assembly in 2003.  Clearly 
visible are the frame, the drift cathode, the field-shaping wires and resistors, 
the Macor pillars, and the Kapton-insulated wires leading to the feedthrough on the flange.  }
  \label{fig:tpc-photos}
  \end{center}
\end{figure}

The final MuCap TPC was constructed and tested at PSI from 2000--2003.
A photograph of the device shortly after its first assembly in 2003 is shown in 
fig.~\ref{fig:tpc-photos}.
A number of novel features were incorporated into the TPC in order to meet 
the experiment's need for high-purity protium gas.  Bakeable, UHV-compatible 
materials were used for all interior surfaces and components.  The materials were 
mostly metals~(Be, Al, Cu, W, Au, stainless steel), metal oxides~(ceramics), 
glass, and Kapton-insulated wires; a small amount of epoxy was used 
to seal the feedthroughs.

The TPC volume had dimensions 
$x\times y\times z = 15\times12\times30$~cm$^3$,
as defined by the rectangular frame of the high-voltage~(HV) drift cathode at the
top of the detector and the rectangular frames of the 
multiwire proportional chamber (MWPC) at the bottom of the detector.  
As shown in 
fig.~\ref{fig:exp-setup}, the TPC was designed to be 
oriented with its long edge parallel to the incoming muon beam.  The frames were 
made from Borofloat glass\footnote{SCHOTT Technical Glass Solutions GmbH,
 07745 Jena, Germany.}
and were held in place by four Macor metal-oxide pillars, one at each corner.  
The MWPC consisted of three separate 
frames---an upper cathode, an anode, and a lower cathode---spaced 3.5~mm 
apart and each having outer dimensions 230~mm~$\times$~380~mm and inner 
dimensions 150~mm $\times$ 300~mm.   The MWPC cathode frames were 5~mm 
thick, while the anode frame was 2.5~mm thick.  Each MWPC cathode frame was
instrumented with 50-$\mu$m-diameter wires oriented parallel to the $z$~(beam) 
direction at 1.0~mm pitch; the wires on the cathode planes were soldered together at their ends in adjacent 
groups of four per soldering pad to produce 35 channels~(``strips'') in total.  
The two cathode wires closest to each edge were larger, with a diameter of
100 $\mu$m, to smooth the field gradient.
Only the signals from the lower MWPC cathode frame were read out; all pads on the upper MWPC cathode were simply connected together to distribute the high voltage.  The anode plane 
had 69 25-$\mu$m-diameter wires oriented in the $x$ direction at 4.0~mm pitch.
An additional three anode wires (two 50-$\mu$m and one 100-$\mu$m in diameter) 
on each end smoothed the field gradient.
The electric field around these larger-diameter wires was not large enough to produce 
gas amplification, so the length of the sensitive region along 
the $z$ direction was effectively reduced to 28~cm.
The wires were made of gold-plated tungsten, with 3\% rhenium.
They were soldered under mechanical tension to Ti-Au-Ni pad structures implanted 
on the glass surfaces.  These soldering pads were developed in close consultation with
the vendor\footnote{IMT Masken und Teilungen, CH-8606 Greifensee, Switzerland.}
to ensure there was no cracking or separation of the pads 
after heating cycles, a problem that had been encountered previously.  The pad layers 
were carefully chosen for their bonding properties: a deposition of a few nm of Ti provided
good adhesion to the glass frames, followed by successive coatings of 
0.5~$\mu$m of gold, 2~$\mu$m of nickel, and 0.1~$\mu$m of gold, where the uppermost gold 
layer was used for soldering to wires.

During operation, voltages of $-29.5$~kV and $-5.5$~kV were typically 
applied to the HV and MWPC cathode planes, respectively, while the anode 
was held at ground potential. This configuration generated a drift field 
of 0.2~kV/mm over the 120~mm vertical distance between the HV frame and the 
upper MWPC cathode frame.  Seven \mbox{1-mm}-diameter copper field wires, wound 
around the Macor pillars at equal vertical spacings and connected in series 
by eight high-impedance 5-G$\rm\Omega$ resistors, created a homogeneous 
electrostatic potential, which ensured field constancy in the sensitive volume 
to within 1\%.  

When a muon came to a stop in the hydrogen gas, it deposited energy
along its path, which ionized surrounding hydrogen atoms.  If this occurred
inside the sensitive volume of the TPC, the positive ions and electrons were
pulled in opposite directions by the detector's uniform electric drift field.  
The ionization electrons drifted downwards at a constant speed 
of $\approx$~5.5~mm/$\mu$s until they reached the MWPC region, where they were
accelerated by its high electric fields to produce an electron avalanche.  
For a 5-kV MWPC field in hydrogen, the resulting signal amplified the initial 
charges by a gain factor of $\approx$~125.

The MWPC anode wires provided $z$-position information ({\em i.e.}, along the muon 
beam direction) while the lower cathode wires provided $x$-position information 
({\em i.e.}, perpendicular to both the muon beam direction and the vertical drift field). 
The $y$ coordinate was calculated from the vertical drift time of the ionization 
charges, as determined by the elapsed time between the muon arrival 
registered in the $\mu$SC detector and the amplification of ionization charges 
in the MWPC.  In this way, the TPC enabled three-dimensional reconstruction 
of the trajectories of muons passing through its sensitive volume.  

The detector had $x$-, $z$-, and $y$-position resolutions of 4.0~mm, 4.0~mm, and 1.1~mm, 
respectively, which were determined by the wire spacing and electronic sampling period.
Magboltz~\cite{Biagi:1999nwa} predicts electron diffusion coefficients of 
65 $\mu$m/$\sqrt{\rm{cm}}$ (longitudinal) and 89 $\mu$m/$\sqrt{\rm{cm}}$ (transverse) in the drift
field, so in principle a substantially better position resolution would 
have been possible.
However, such fine spatial resolution was not needed for this experiment.
It should be mentioned that the manner in which the TPC entangles 
time and space in the $y$ dimension can produce subtle effects which must be 
carefully addressed in a high-precision experiment such as MuCap.  These 
issues are discussed in detail in sect.~\ref{sssec:decay-spectrum}.

The TPC was mounted on an aluminum fork affixed to the stainless steel flange 
at the downstream end of the pressure vessel.  The flange contained electrical 
feedthroughs for the high voltage and signal lines.  All of the TPC's signal 
preamplifiers were attached to the outside of the flange and were mounted as close 
to it as possible to minimize noise pickup and wire capacitances.  
The flange was attached at its center to a cantilevered tube that
both supported the pressure vessel assembly and functioned as a high-vacuum 
conduit during evacuation
and baking operations before the protium gas was loaded.

The TPC sat in the center of the cylindrical pressure vessel, which was made of 
4-mm-thick Anti\-corodal-110 aluminum and had length 600~mm and inner 
diameter 282~mm.  The stainless steel flange on the upstream end of the vessel 
contained the beam entrance window, a beryllium half-sphere of thickness 0.5~mm 
and radius 35~mm. The beryllium window and the pressure vessel's bare 
aluminum wall were made as thin as safely possible in order to minimize 
scattering of incoming muons and outgoing decay electrons, respectively.

While the MuCap setup was designed to detect negative muons and their decay electrons, it was essential that the apparatus also be able to measure the lifetime of positive muons as a control experiment.
Positive muons remain polarized 
after they stop, and the distribution of the decay positrons is 
peaked in the direction of the muon's spin.  
Rather than allow the muon spins to precess in an unknown and possibly variable 
ambient magnetic field, it was better to impose a known magnetic field to rotate 
the spins in a controlled way.
A dual saddle-coil magnet, wound around the pressure vessel from 6.4-mm-diameter
aluminum tubes carrying a current of 125~A, provided a 5~mT field,
uniform at the level of $\sim$~10\% over the sensitive volume.

In addition to the TPC that was used for the physics measurement, a second
one of the same general design was constructed and tested as a spare.  
It was tested with a $^{90}$Sr beta source up to 6~kV~MWPC high voltage, 
and it remained stable for one week at a working voltage of 5.5~kV.

\subsection{Electronics}
\label{ssec:electronics}

\begin{figure*}[tp]
  \begin{center}
  \includegraphics[width=0.7\textwidth]{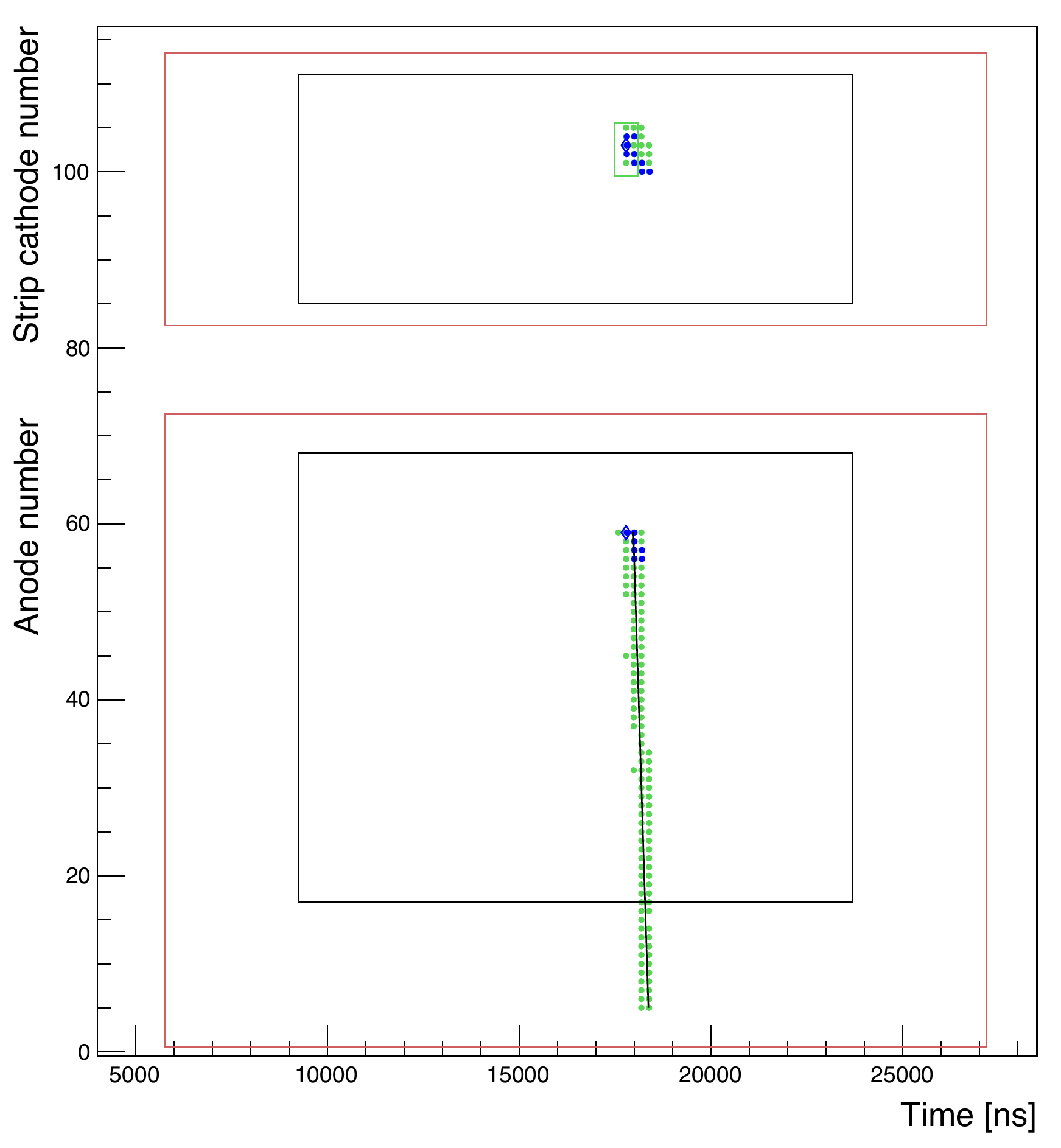}
  \caption{A typical beam muon stop in the TPC, as presented in the MuCap event 
  display.  Each incoming muon deposited energy along its path in a characteristic 
  manner, indicated by the pixel colors: the green EL (Energy Low) pixels correspond 
  to lower-energy ionization along the muon's trajectory, while the blue EH (Energy High)
  pixels correspond to the large Bragg peak at the muon's stopping point.  A similar 
  signature was produced in the cathodes, shown at the top of the display.
  The boundaries of the TPC's sensitive volume are indicated by the red boxes; the fiducial volume of muon stops accepted for analysis is drawn as a black box.  The straight-line fit to the track and the stopping point (diamond shape) determined by the analysis software are drawn on top of the track.
  The edges of the boxes in the $y$/time dimension were established by the muon arrival time 
  recorded in the $\mu$SC detector.
  }
  \label{fig:evdisplay-mustop}
  \end{center}
\end{figure*}

The amplified TPC signals were processed by VME64 time-to-digital-con\-verter~(TDC) 
modules built for the MuCap experiment.  These PSI-designed electronics
modules, model TDC400, recorded time-stamped hit patterns (48 bits for data, 
16 bits for time) from the MWPC wires at a rate of 5~MHz and at three signal 
thresholds, which served to reduce data volume.  The optimal settings for the three 
discriminated signal levels---designated Energy Low~(EL), Energy High~(EH), 
and Energy Very High~(EVH)---varied according to the hydrogen gas density and the 
TPC's operating voltage, so they were tuned by eye using a graphical event display 
(fig.~\ref{fig:evdisplay-mustop}) and trend plots prior to data taking.  Whenever an 
energy threshold was triggered, all lower-energy thresholds were also triggered.  A TDC400 module was typically used to instrument each 16-channel sector of the TPC 
anodes, the 48 data bits being used for the EL, EH, and EVH signals from that sector.

The TPC operated in proportional mode, so its signals were proportional in voltage
to the energy deposited per unit length by ionizing particles.  The EL threshold was
set just above the noise so that each incoming muon generated a trail of EL pixels 
along its ionizing track.  The EH threshold was tuned slightly higher, so that the track 
terminated in a cluster of EH pixels at the muon's stopping point where there was a 
large Bragg peak in the energy deposition.  The EVH threshold was generally set just 
above the 250~keV high-energy tail of the Bragg peak in the muon stopping 
distribution to render the detector sensitive to signals from exotic high-energy 
processes---namely, nuclear muon capture by $Z>1$ impurities, which typically 
produced recoil nuclei in the 200--800~keV energy range.  During 2006 data taking 
typical threshold settings were in the vicinity of EL=6~keV, EH=75~keV, 
and 
EVH=340~keV.

During the 2004 experimental run, the central 16-wire TPC sector (anodes 33--48) 
was also instrumented with 8-bit flash ADC~(FADC) VME modules (Struck DL401).
The FADCs provided analog information on the TPC anode 
	signals, albeit at a limited data rate compared to the TDC400s, as the FADCs were 
	only read out on infrequent occasions when the $Z>1$ impurity capture trigger logic 
	fired.  
	In subsequent years, all of the TPC channels were instrumented with FADC modules
	custom designed for the experiment.  These modules, built around the Maxim 
	MAX1213 device, transmitted data directly over an Ethernet interface.  In comparison 
	to the previous setup, they provided full channel coverage, 12-bit resolution, and a 
	higher allowed trigger rate.

	\subsection{Gas system}
	\label{ssec:gas-system}

\begin{figure*}[tp]
	  \begin{center}
	  \includegraphics[scale=0.50]{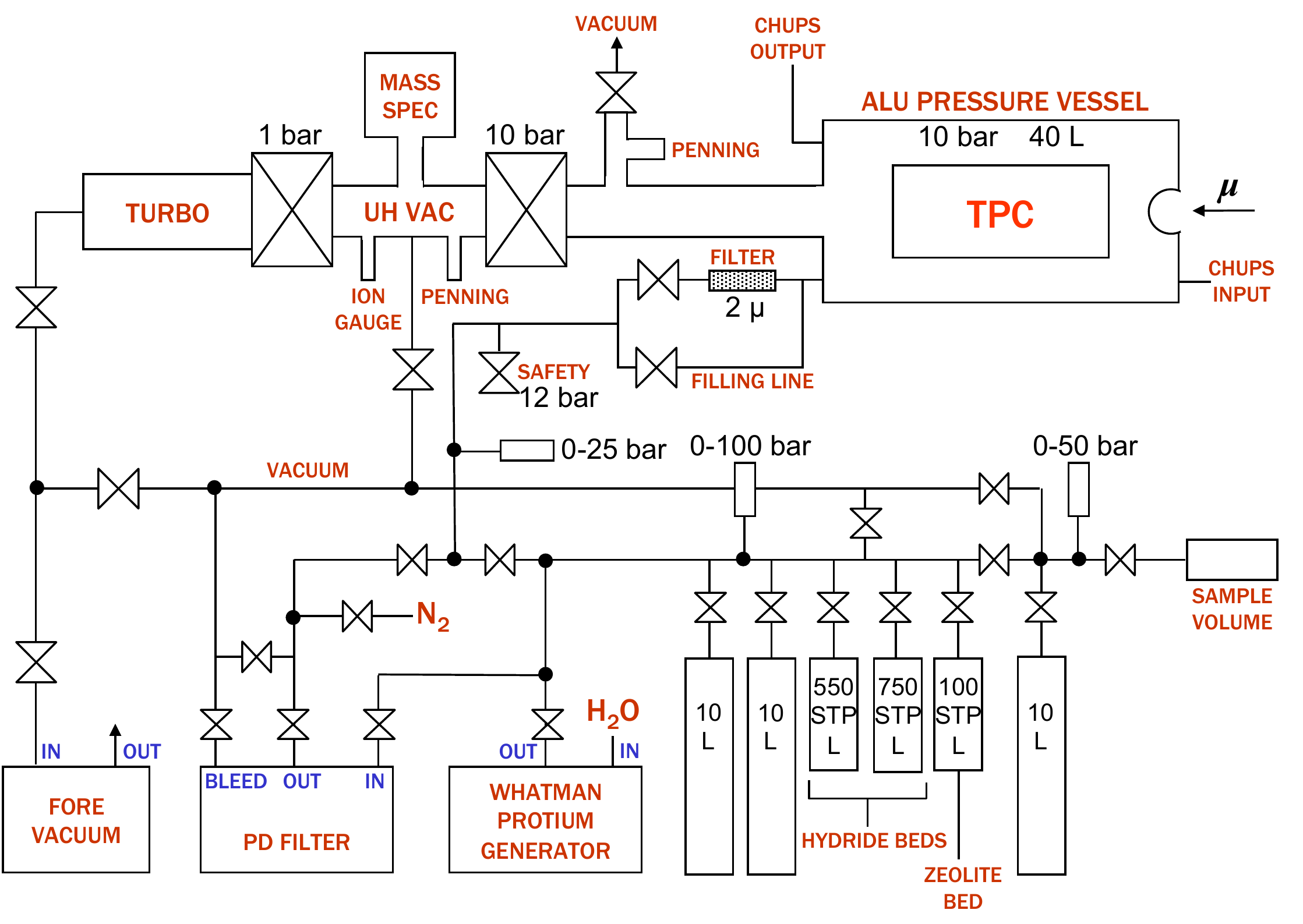}
	  \caption{Schematic diagram of the TPC's high-vacuum and gas-handling 
	  system.
	  }
	  \label{fig:gas-system}
	  \end{center}
	\end{figure*}

	A combination high-vacuum/gas-handling system (fig.~\ref{fig:gas-system}) was 
	used to store the ultra-pure protium and to vacuum pump and then fill the 
	pressure vessel with protium gas at the start of every physics run.

	Prior to each run, the pressure vessel (with TPC installed inside) was 
	high-vacuum pumped while being baked to 110--130$^\circ$C
	to remove impurities.  
	The low temperature-expansion coefficient of the TPC's glass frames prevented 
	excess tension on the wires during these heating cycles, and no cracking or 
	separation of the custom wire-soldering pads was observed after repeated 
	heatings.

	After a baking cycle, a residual pressure below $10^{-7}$~mbar 
	could be maintained in the pressure vessel by high-vacuum pumping,
	using a 200~L/s turbomolecular pump in combination with oil-free forevacuum 
	pumping.  The main 
	pumping of the pressure vessel was conducted through the 600-mm-long, 
	71~mm-inner-diameter support tube attached to the center of the downstream flange.  
	All vacuum connections were sealed with metallic gaskets except for the two end 
	flanges of the pressure vessel, in which Viton O-rings were used.  A purely metallic 
	valve with a 70~mm aperture isolated the portion of the system under vacuum from the high-pressure part.  
	The vacuum was monitored by Penning and Pirani gauges, and a quadrupole mass 
	spectrometer was used to monitor the composition of the residual gas.

	The protium gas handling system consisted exclusively of stainless steel components 
	that could be baked out in parallel with the pressure vessel and could tolerate 
	pressures up to 60~bar.  The system contained the following gas-storage devices:
	\begin{itemize}
	\item Hydride storage beds: Each of these 2-liter containers stored protium gas
	bound as a hydride at residual pressures of 1--10~bar, depending on the filling 
	state and temperature.  Typical storage capacities were 500--800 STP-liters per container.
	\item A zeolite bed: At liquid nitrogen temperature, this 1-liter-volume bed 
	adsorbed up to 120~STP-liters of protium.  At room temperature the protium gas 
	was desorbed and then pressurized up to~$\sim$~100~bar so that it could be
	filled into stainless steel bottles.
	\item Stainless steel bottles: Each of these 10-liter bottles could be filled with 
	pressurized gas and had a maximum storage capacity of 600~STP-liters. Smaller 
	one-liter volumes were used for gas mixing and sampling duties.
	\end{itemize}

	The protium gas was produced via electrolysis of commercially purchased 
	protium water\footnote{Ontario Power Generation, Ontario, Canada} using a
	Whatman generator.  The gas's 
	residual deuterium level of 1.44~ppm, though quite low, nevertheless 
	necessitated a significant correction to the data collected during the 
	experiment's first physics run in 2004~\cite{Andreev:2007wg}.  An isotope separation 
	column was therefore built in 2005 to further reduce the deuterium content
	of the hydrogen gas for subsequent measurements.  The separation column 
	employs hydrogen liquefaction and evaporation cycles to purify the protium 
	by exploiting the different evaporation rates of hydrogen isotopes; a complete 
	description of the system is given in ref.~\cite{Alekseev:2006}.  A sample of 
	ultra-depleted protium gas produced by the column and used in the TPC
	was analyzed by the ETH AMS facility~\cite{Synal:2006} and found to contain less 
	than 6~ppb deuterium.  Both the isotopic purity and the sensitivity of
        the analysis are unprecedented, representing significant advances 
        in the state of the art.

	During storage the protium accumulated $Z>1$ impurities due to outgassing from its 
	containers, so the gas had to be cleaned prior to its injection into the 
	pressure vessel through the single 6-mm-OD/4-mm-ID stainless steel tube.  
	Initial cleaning was accomplished by passing the protium through a palladium filter, 
	which is effective at removing non-hydrogen elements.  
	Although the resulting hydrogen gas was of ultra-high purity, continued 
	cleaning was required because the vessel's interior and the TPC continually 
	outgas impurities.  Even after the vessel 
	and TPC had been heated and pumped for several weeks the system still 
	exhibited an impurity outgassing rate of~$\sim$~1 ppm/week; analysis of gas 
	samples revealed that nitrogen and water vapor were the primary 
	impurities~\cite{Banks:PhD,Clayton:PhD}.  To overcome this problem, the 
	Circulating Hydrogen Ultra-high Purification System~(CHUPS) was 
	operated during data taking.  
	This system---which worked on the basis of thermodynamic 
	adsorption and desorption cycles using activated carbon---continuously 
	circulated the protium gas through the pressure vessel via dedicated lines 
	(consisting of an inlet in the upstream flange and an outlet in the downstream 
	flange) and scrubbed the protium of impurities by pushing it through zeolite filters 
	in a liquid nitrogen environment.  A detailed description of CHUPS and its 
	performance can be found in ref.~\cite{Ganzha:2007uk}.

Several ports were installed to allow gas from inside the pressure vessel
	to be sampled (up to~$\sim$~10~STP-liters) from the single connecting tube and 
	stored in bottles for later offline analysis.  The impurity content of the gas was 
	analyzed using gas chromatography, which is capable of detecting small 
	admixtures of light elements (N$_2$, O$_2$) down to 10~ppb.  The humidity
	content was measured online using a Pura hygrometer.\footnote{Kahn Instruments, 
	Wethersfield, CT, USA}

	In the 2004 physics run, the gas-handling setup was also used to take samples
	of the gas inside the pressure vessel and to inject precise admixtures of impurity
	gases for calibration studies, using ports connected to the single stainless steel 
	tube between the gas handling system and the pressure vessel.  In 
	subsequent runs these tasks were performed using ports within the separate 
	CHUPS system, which allowed for cleaner operations.

	\section{Performance}
	\label{sec:performance}

\subsection{Wire chamber properties}

\subsubsection{Gas amplification}

\label{sec:gain}

A wire chamber with a high gas gain separates small signals more
effectively from the noise.  
We characterized the gas gain in several prototype chambers before building the MuCap TPC.
Measurements of the gain of Gatchina-built MWPC prototypes
were reported in~\cite{Maev:2001yx}.  One of these chambers had a 
similar geometry to the corresponding region of the MuCap TPC.  
The measurements were made using an alpha source, so the recombination effects 
significantly reduced the charge collection and therefore the apparent gain.
Consequently, these results were not directly comparable to the conditions 
of the MuCap experiment.  

\begin{figure}[tp]
  \begin{center}
  \includegraphics[width=\columnwidth]{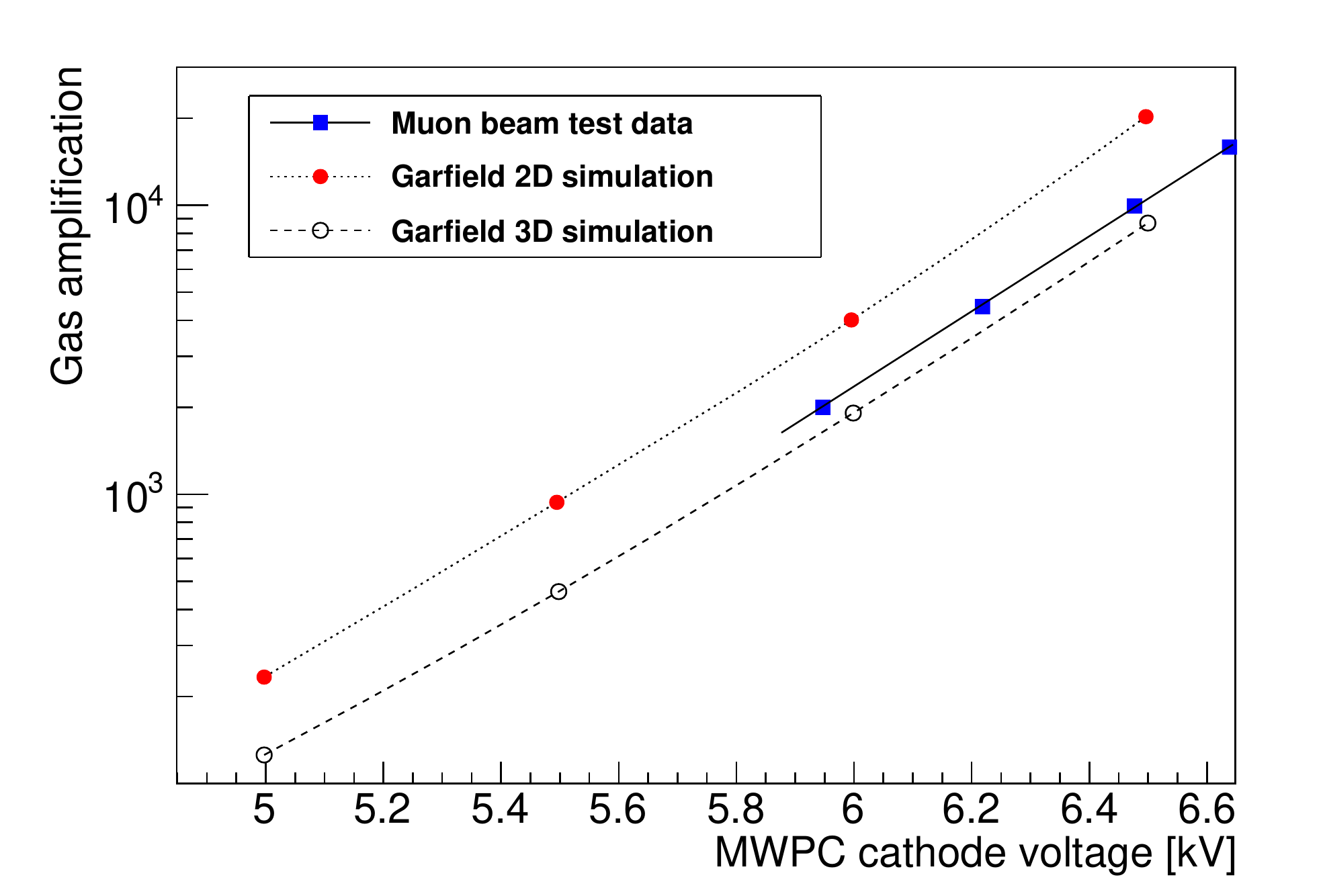}
  \caption{Gas amplifications from measurements of electrical current in a 
prototype MuCap MWPC exposed to beamline muons (squares), as well as from 
Garfield simulations for an approximate 2D MWPC geometry (closed circles) 
and a more realistic 3D MWPC geometry (open circles).}
  \label{fig:gainVsHV}
  \end{center}
\end{figure}

However, a PSI-built MWPC prototype with the same geometry
as the final detector was also tested with a 50~kHz beam of 32~MeV/$c$ 
muons at PSI.  The energy deposited by each muon passing through the 7~mm 
depth of the MWPC region was calculated
to be 42~keV; since each ion pair requires 37~eV in hydrogen, the primary 
ionization current was $9.1 \times 10^{-12}$~A.  The ratio of the cathode power supply 
current to this value of ionization current was defined as the gas 
amplification.  These values are plotted in fig.~\ref{fig:gainVsHV}, where
they are compared with 
the results of Garfield calculations, both in the
traditional two-dimensional approximation where the cathodes and anodes
are treated as parallel, and in a realistic three-dimensional geometry 
calculated by neBEM~\cite{Majumdar:2006jf}, where the anode and cathode 
directions are perpendicular.

\begin{figure}[tp]
  \begin{center}
  \includegraphics[width=\columnwidth]{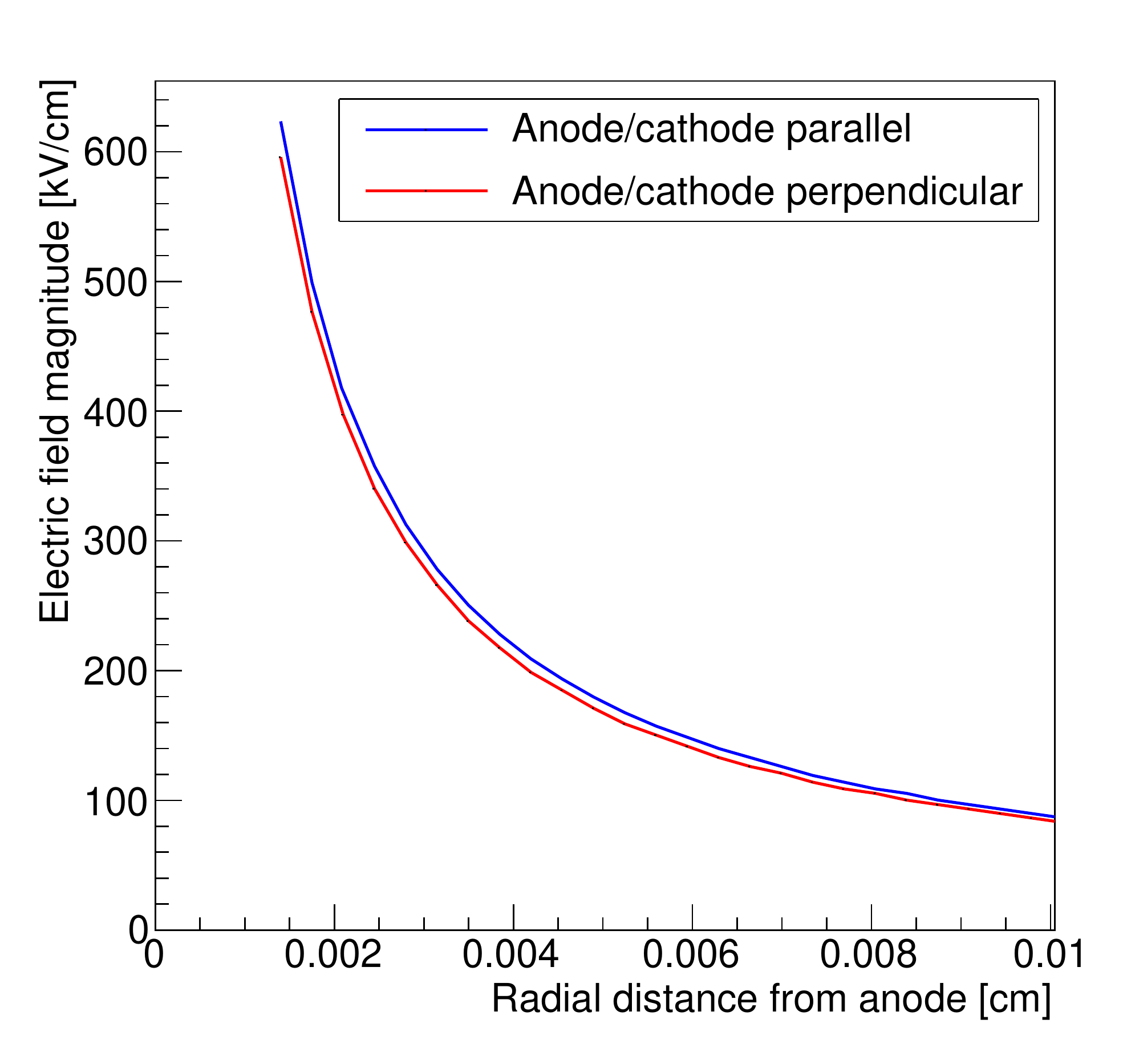}
  \caption{Electric field magnitude vs.\ distance from anode wire,
  calculated both with a realistic three-dimensional geometry and
  with the two-dimensional approximation of parallel wires.
  An MWPC cathode high voltage of $-6$~kV is assumed here.
  Because the Townsend coefficient increases dramatically with the field
  strength, the seemingly small difference can account for a factor of $\sim$~2 
  in the calculated gas amplification.}
  \label{fig:EvsR}
  \end{center}
\end{figure}

The experimental results 
agreed within $\sim$~15\% with the three-dimensional calculation, which 
gave a result that is a factor of $\sim$~2 smaller than the two-dimensional
calculation.  Clearly, in this case, the relative orientation of the 
anode and cathode wires (parallel or perpendicular)
has a significant effect.  Because the Townsend coefficient 
rises very sharply with the electric field magnitude $E$ as described by 
eq.~\ref{eq:townsend}, a difference in $E$ near the anode of a few 
percent was sufficient to account for this 
discrepancy.  This radial dependence of $E$ with respect to the center of
the anode is plotted in fig.~\ref{fig:EvsR}.

\subsubsection{Ion transparency}

In high-rate experiments with particle fluxes at the level of $10^5$~Hz/cm$^2$
or more, it is essential to minimize the transparency of the amplification 
stage of a gaseous detector to the ions produced in the avalanche.  
These ions, if released back into the detection volume (ion feedback),
can distort the fields there by their space-charge effect.
The question of ion feedback motivates the application of
technologies such as GEM~\cite{Sauli1997531} and
Micromegas~\cite{Giomataris199629} in many detector systems.

\begin{figure}[tp]
  \begin{center}
  \includegraphics[width=\columnwidth]{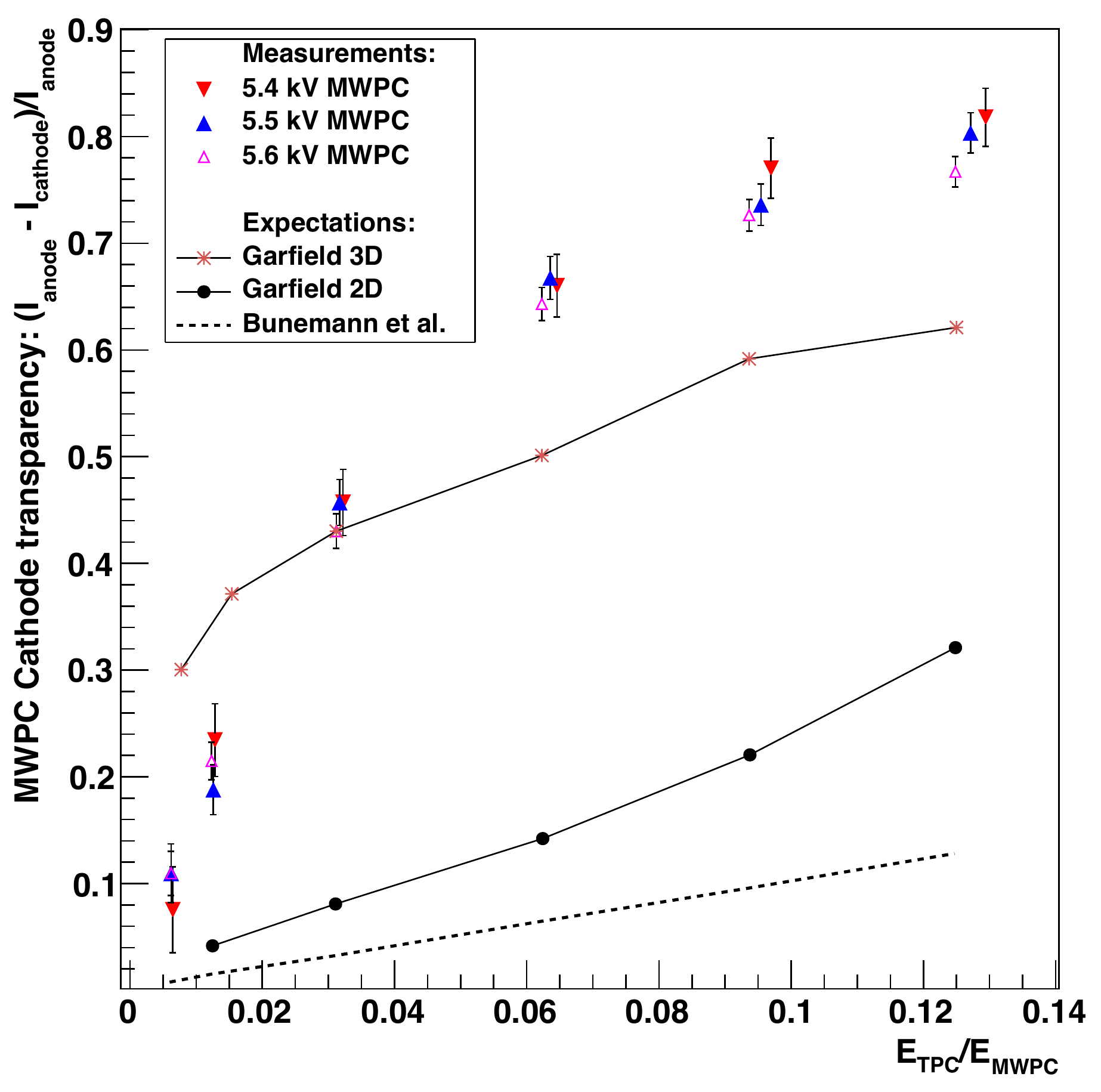}
  \caption{Measurements of the ion transparency of the MWPC upper 
  cathode versus the ratio of the electric fields in the drift and 
  MWPC regions.  The transparency significantly exceeded the predictions of 
  simple models and was explained by ``channeling'' of field lines in three 
  dimensions, as discussed in the text.}
  \label{fig:transparent}
  \end{center}
\end{figure}

MuCap was not a high-rate experiment in this sense; the incident muon beam rate 
was only $\sim~3 \times 10^4$~Hz and the gas gain was low,
so ion feedback was not a major design 
consideration.  It was nevertheless interesting to investigate the transparency
of the MWPC cathode plane in order to understand whether ion feedback
may have contributed to the instability of the TPC at higher MWPC voltages,
and also to be able to interpret the MWPC cathode current measured 
during the physics data taking. 
fig.~\ref{fig:transparent} shows measurements of 
the ion transparency, made using a $^{90}$Sr source, in which the difference between 
the MWPC anode and cathode currents was examined as the drift field in 
the TPC was varied.  This current imbalance arose because ions that passed 
through the upper MWPC cathode were collected instead at the drift cathode.  

The measured transparency was much larger than naively expected from the 
calculation in~\cite{Bunemann}, which suggests that the transparency should 
simply equal the ratio of the electric fields in the drift and MWPC regions, 
$E_{TPC}/E_{MWPC}$.  We note that this calculation is effectively the 
reciprocal of the problem of electron transmission from the TPC into the MWPC, 
which is essentially 100\% at the MuCap operating conditions, in good agreement 
with the predictions of~\cite{Bunemann}.
However, this relation was derived for electrons drifting 
in an ionization chamber with a Frisch grid and a planar anode, which does not 
directly apply to
the MuCap TPC.  It assumes that the anode wire pitch was much less than the 
half-gap between anode and cathode planes.  This result would not, 
in principle, be expected to apply to the MuCap TPC, because of its larger wire 
spacing.  The calculation also assumes that the anode and cathode wires 
are parallel to each other. Also, it describes only the transparency to 
electric field lines, and not to drifting ions.  

A two-dimensional Garfield simulation that traced the field 
lines predicted a somewhat larger transparency, though still not large enough 
to match the experimental data.
The three-dimensional Garfield simulation revealed the origin of this 
large ion transparency.  The cathodes focused each incident electron onto an $x$ 
location along the anode that was equidistant from the adjacent cathodes.  
The avalanche developed primarily within 50~$\mu$m of the anode wire, so it 
remained localized near this $x$ coordinate.  Ions therefore generally 
originated from locations where the electric field lines led out of the MWPC.
This ``channeling'' was seen in the simulation by tracking an electron from 
the drift region to an anode, then randomly perturbing each component of the 
track endpoint by $\pm$50~$\mu$m as an approximation to the avalanche volume, 
and then tracking an ion from that location.  This crude model was used because 
full avalanche tracking in the detailed field map was too computationally 
intensive, but nevertheless it described the scale of the observed transparency 
reasonably well.

\subsubsection{Drift speed}

The TREAD group~\cite{Chapin:1984uk} gave a parameterization of TPC
drift speed in terms of electric field and pressure.
The MuCap experience suggests that this formula provides a useful
first approximation.
It predicts $5.88 \pm 0.12$~mm/$\mu$s for the
MuCap conditions, and this is the same value calculated by Garfield.
This number agrees only at the level of 3$\sigma$
with the speed of 5.5~mm/$\mu$s observed from the 
distribution of drift times, which does match the experience of
the $\mu$CF experiment~\cite{muCF}.

	\subsection{Experimental operation}
	\label{ssec:exp-operation}

	The MWPC part of the TPC was designed to operate at voltages up to 6.5~kV, at 
	which it is in principle capable of detecting not only the tracks of incoming 
	muons but also the tracks of outgoing minimum-ionizing decay electrons.  
	In practice, this design voltage was never sustained because 
	excess currents and sparks typically 
	began to appear in the range 5--6~kV, leading to voltage trips.
        This behavior was unexpected, since the prototype TPC was able to 
        operate at 6.4~kV, successfully detecting minimum-ionizing 
        and Auger electron signals~\cite{Maev:2001,Maev:2001yx}.

	At the outset of the 2004 data taking run, the TPC achieved a maximum stable 
	operating voltage of 5.2~kV.  Over the ensuing weeks, the maximum voltage 
	gradually deteriorated to 4.8~kV; most of the 2004 data was collected at 5.0~kV.  
	The exact reasons for this behavior remain unclear, though it is hypothesized 
	that dust accumulation on the MWPC wires could be responsible.  A 
	\mbox{1-mm}-thick Al plate was installed underneath the MWPC prior to the 2004 run 
	in an attempt to prevent dust from collecting on its wires.

	For the 2006 run, the maximum stable operating voltage 
	was increased to 5.45~kV.
	This was accomplished through improvements in the procedures for cleaning 
	the TPC between experimental runs, and by conducting more cautious training 
	of the TPC in the months leading up to the runs ({\em e.g.}, the operating voltage was
	ramped up very slowly).  The half-gaps between the MWPC planes were 
	increased from 3.5~mm to 3.64~mm; this change required the high voltage 
	be increased by 90 V to achieve the same gain as before. Unfortunately, stable 
	operation 
	at the increased operating voltage was not possible in 2007, and it had to be 
	lowered again.  These running conditions and the 
	associated gas amplification factors predicted by Garfield are 
	summarized in Table~\ref{tab:runParameters}.  The validation of the
	Garfield simulations with experimental data is described in sect.~\ref{sec:gain}.

	\begin{table}
	\begin{center}
	\begin{tabular}{llll}
	\hline
	Year & 2004 & 2006 & 2007 \\
	\hline
	Cathode voltage (kV) & 5.0 & 5.45 & 5.1 \\
	Half-gap (mm) & 3.5 & 3.64 & 3.64 \\
	MWPC gas gain & 125 & 320 & 132 \\
	\hline
	\end{tabular}
	\caption{Typical operating conditions of the MuCap TPC in its three 
	major data-taking runs.  The gas gains were calculated with the
	Garfield program.}
	\label{tab:runParameters}
	\end{center}
	\end{table}

\begin{figure*}[tp]
	  \begin{center}
	  \includegraphics[width=\textwidth]{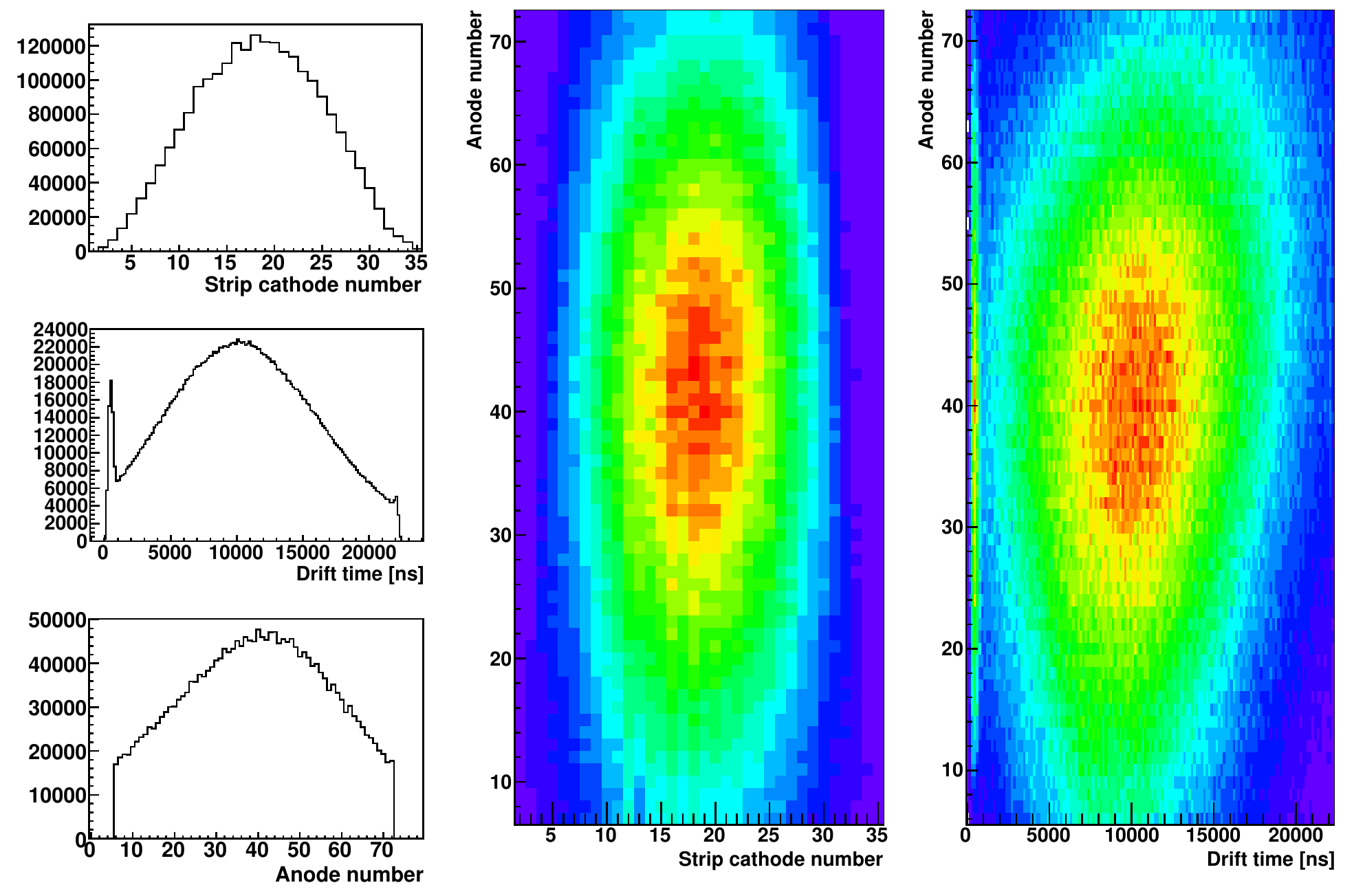}
	  \caption{Muon stopping distributions from a typical run in 2004. 
	  In the left column, $x$, $y$, and $z$ projections are shown; in the
	  right columns are two-dimensional $xz$ and $yz$ spectra.  
          The strip cathode numbers increase in $-x$, drift time increases
          in $+y$, and anode numbers increase in $+z$.
	  The sharp peak in the drift time ($y$) distribution at small times
	  corresponds to muons that passed through the MWPC region, causing 
          ionization directly there.}
	  \label{fig:TPC-online-display}
	  \end{center}
	\end{figure*}

	Prior to the start of every run period---and also whenever the TPC's maximum 
	stable operating voltage changed during a run---the three signal energy 
	thresholds (EL, EH, EVH) were manually tuned until incoming muon tracks 
	exhibited the correct appearance in the graphical event display (see 
	fig.~\ref{fig:evdisplay-mustop}).  During data taking, operators continuously
	monitored the stopping distribution of beam muons in the TPC using an 
	online display~(fig.~\ref{fig:TPC-online-display}) which relied on a simplified 
	version of the analysis software to process the incoming data in real time.  
	Depending on the beam tune, approximately 55\%--65\% of the muons passing 
	through the $\mu$SC detector stopped inside the sensitive volume of the TPC.  

	\subsection{Data analysis}
	\label{ssec:data-analysis}

	\subsubsection{Identification of muon tracks}
	\label{sssec:mutrack-ID}

	A good muon stop in the TPC characteristically appeared in the discriminated 
	anode data as a nearly continuous trail of EL pixels terminating in one or more 
	EH pixels (see fig.~\ref{fig:evdisplay-mustop} and the accompanying text). 
	A similar signature appeared simultaneously in the cathode data, though 
	this direction was generally more complex to interpret because it was parallel 
	to the muon velocity; also, the anodes collected the drift charges while the 
	cathodes received only a pickup signal.  Our basic approach to identifying muon 
	stop candidates in the TPC, therefore, was
	to search for clusters of EH pixels in the anode data and then inspect the 
	surrounding anode and cathode data for evidence of an associated EL track.

	To facilitate track analysis, the raw TPC data was first placed into a time- and 
	position-ordered array consisting of three stacked layers, where each layer 
	contained hit information from one of the discriminated energy 
	thresholds: EL, EH, and EVH.  The track-finding algorithm began by searching 
	the anode EH data for clusters of one or more nearly contiguous pixels.  
	Once the boundaries of an EH cluster had been determined, the software 
	searched in the upstream and downstream directions for an attached track by
	following any trails of anode EL pixels leading away from the EH cluster.  
	Small gaps between pixels were allowed, and the algorithm was capable of pursuing 
	any number of branches in the trail.  In this way, the software ascertained the length 
	and extent of any tracks connected to the EH cluster.  Lastly, the algorithm 
	searched for cathode hits coincident with the anode EH cluster (to get the 
	$x$~position of the muon stop) and with any tracks downstream of the EH cluster 
	(to identify muons that escaped out of the side of the TPC).

	Though simple, this algorithm was consistent and reliable, and it enabled 
	analysis and correction of systematic biases.  The analysis of data from 
	the 2004 run used only this approach of counting pixels in EH clusters and 
	connected EL tracks, requiring that they be within the fiducial volume; in that 
	analysis, a correction was applied to account for muons scattered 
	through large angles out of the TPC.  This correction was avoided in the 
	analysis of data from the 2006--2007 runs by imposing additional requirements 
	on the muon track.
	Because the maximum range of the proton following a $\mu - p$ scattering event
	was $\sim$~1~mm, it was able to trigger at most one EH pixel.  Consequently, a 
	minimum of two pixels was required to consider an EH cluster as a muon stop.
	Each track was also required to fit acceptably to either a single straight line 
	or a two-segment bent line that would correspond to a small-angle
	forward scatter.   Cuts were also made requiring a minimum track length
	and enforcing a maximum distance between the stop location and the track endpoint.
	While these additional requirements effectively removed the large-angle
	scattering events, they also increased the sensitivity of the muon stop 
	definition to interference from minimum-ionizing Michel electron tracks
	if they crossed near the stopping point, requiring a new correction for that 
	effect.

	\subsubsection{Fiducial volume cuts}
	\label{sssec:fid-cuts}

	To select the desirable muons that stopped inside the TPC's fiducial volume, a 
	series of cuts were applied to the track information collected for each candidate.  
	In the $x$~dimension there had to be cathode hits (EL or EH) coincident in time 
	with the anode EH pixels corresponding to the muon's ostensible stopping point.  
	Tracks with cathode hits near the edges of the sensitive volume were cut in order 
	to discard muons that might have escaped through the sides of the TPC.  Similar 
	considerations applied to the $z$~dimension, though in addition there had to
	be a trail of anode EL hits leading up to the EH pixels of the muon stop.

	Fiducial cuts in the $y$~dimension required special care, for the following 
	reason: Although outgoing decay electrons did not generate tracks in the TPC, 
	we discovered that they occasionally deposited sufficient energy to trigger the 
	EL threshold, and that this effect was correlated with the direction of emission.  
	As a result, cuts based on the presence of extraneous EL pixels in the immediate 
	vicinity of the muon's stopping point skewed the observed decay time spectrum in 
	both a time- and space-dependent manner---in particular, they generated a 
	nonstatistical nonuniformity in the decay rates measured by the 16 segments 
	of the surrounding eSC detector.  To avoid this distortion, we performed a 
	``double-walled'' cut in which all pixels associated with a muon track had 
	to be contained inside a pair of outer boundaries (to ensure the muon did not leave 
	the TPC) while the EH pixels corresponding to the Bragg stop had to lie inside 
	a pair of inner boundaries.  The modest spatial separation between these two 
	sets of boundaries accomplished the required fiducial cut while effectively 
	avoiding both systematic distortions and excessive losses in statistics.

	The fiducial cut boundaries in all three dimensions are illustrated in 
	fig.~\ref{fig:evdisplay-mustop}.
	More detailed explanations of the muon track topologies and fiducial cuts 
	can be found in refs.~\cite{Banks:PhD,Clayton:PhD,Kiburg:PhD}.

	\subsubsection{Decay time spectrum}
	\label{sssec:decay-spectrum}

	The immediate goal of the MuCap experiment was to determine the effective 
	$\mu^-$ disappearance rate in hydrogen from the measured time spectra 
	of decay electrons.  These spectra were constructed by histogramming the 
	elapsed times between muon arrivals and electron emissions, 
	$\Delta t = t_e - t_\mu$, as established by the $\mu$SC and eSC detectors.

	The TPC was used to select the $\mu$SC arrival times that corresponded to good muon 
	stops in the hydrogen gas.  However, the TPC itself relied on the $\mu$SC to 
	determine the drift times (and hence $y$~positions) of muon tracks.  This 
	interdependency led to problems when multiple tracks were present in the TPC 
	at the same time: due to the entanglement of time and position in the $y$~dimension, 
	in such cases it was impossible to unambiguously match $\mu$SC arrival times with 
	muon tracks in the TPC.  If left untreated, this ambiguity would have
	allowed some $\mu$SC times to be approved by uncorrelated TPC tracks, 
	thereby generating two ``wrong-electron'' components in the 
	accidental background of the decay time 
	spectrum: an innocuous uniform background, and a wide hump whose peak 
	would lie underneath the decay time spectrum.

\begin{figure*}[tp]
	  \begin{center}
	  \includegraphics[scale=0.65]{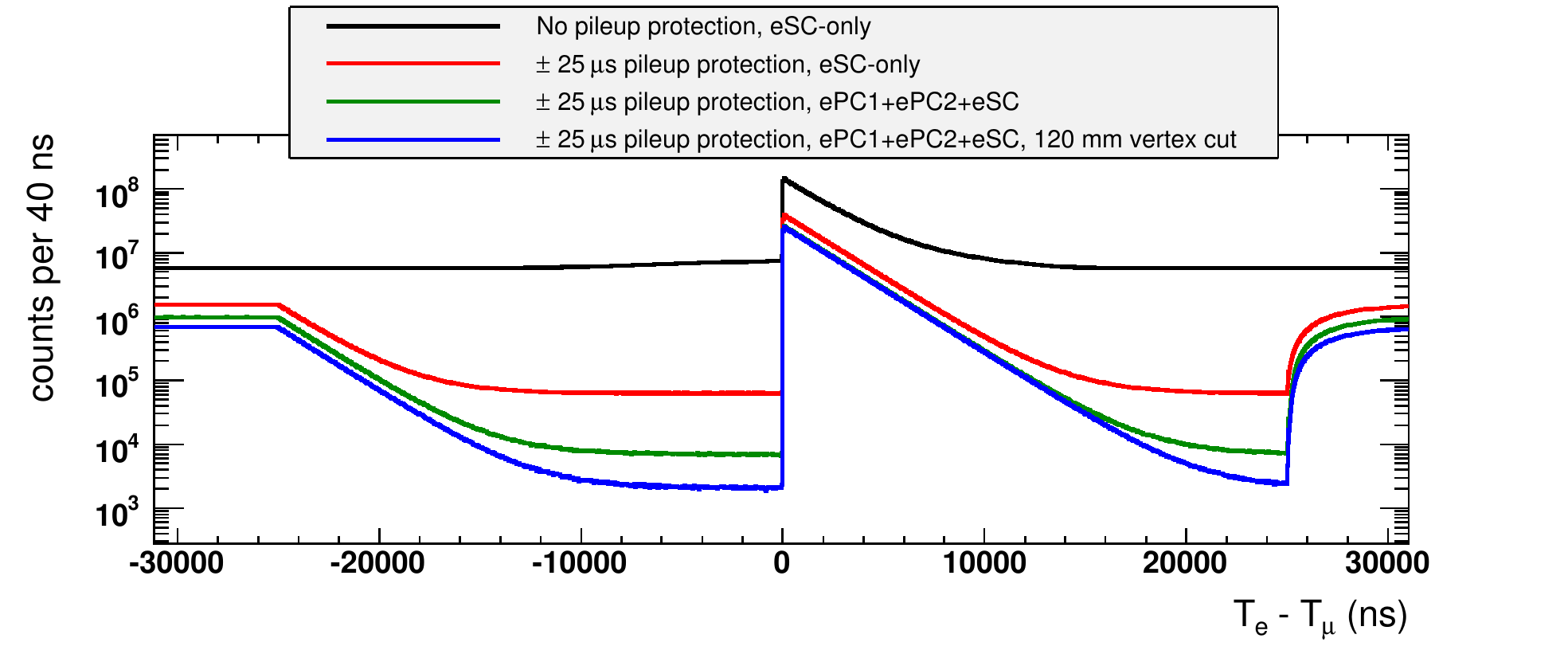}
	  \caption{Muon decay time spectra before and after performing $\pm25~\mu$s 
	  pileup protection on muon arrivals, as well as various electron 
          tracking and $\mu-e$ vertex cuts, in the 2004 data.
	  The procedure effectively removes the accidental background 
	  beneath the decay exponential; the shape of the remaining background 
	  is a convolution of the exponential with a negative square pulse.  
	  Note that the small hump near $T_e - T_\mu = 0$ in
	  the accidental background---the primary motivation for pileup protection---
	  is slightly visible in the spectrum without pileup protection.  }
	  \label{fig:decay-time-spectrum}
	  \end{center}
	\end{figure*}

	To avoid the latter, unwanted distortion in the accidental background, only muons 
	separated in time by $\pm$25~$\mu$s from other muon arrivals were accepted.  
	The pileup protection interval was deliberately chosen to be slightly larger than the 
	maximum TPC drift time of $\approx23.5~\mu$s to ensure there was only one muon 
	track in the TPC in for the full drift period following a muon arrival.
 In the 2004 run
	this condition cut the usable statistics by $\sim$~68\%, but it eliminated the possibility 
	for ambiguity when associating $\mu$SC times with TPC tracks, and it dramatically 
	improved the signal-to-background ratio.  The resulting effect on the decay time 
	spectrum is shown in fig.~\ref{fig:decay-time-spectrum}.

	A muon-on-demand kicker apparatus~\cite{Barnes}
        was installed in the PSI $\pi$E3 beamline 
	in preparation for the 2005 running period.  This device enabled individual 
	muons to be gated into the detector at controlled times, largely
        avoiding the 2004 pileup losses and increasing the good data collection
       rate by a factor of $\sim$~3.

	Further suppression of the accidental background in the decay time spectra
	was achieved by performing a vertex cut on the impact parameter between 
	each decay electron's trajectory and its parent muon's stopping point,
        as shown in fig.~\ref{fig:decay-time-spectrum}.  It is also
	worth noting that the concentration of deuterium in the protium gas and its 
	effects could be determined by studying how the decay time spectra vary with
	the size of the vertex cut~\cite{Clayton:PhD}.

	\subsubsection{$Z>1$ hydrogen gas impurities}
	\label{sssec:impurities}

	The TPC also functioned as an in situ 
	monitor of the levels of unwanted $Z>1$ elements in the purified
	hydrogen gas.  This capability proved invaluable, as it enabled 
	us to track contamination levels and evaluate the performance of 
	the hydrogen cleaning and circulation system~(CHUPS) in near-real time.  
	The TPC data was also used in the offline analysis to perform a 
	zero-extrapolation correction for the effects of residual impurities on the 
	decay time spectrum.

\begin{figure*}[tp]
  \begin{center}
  \includegraphics[width=0.7\textwidth]{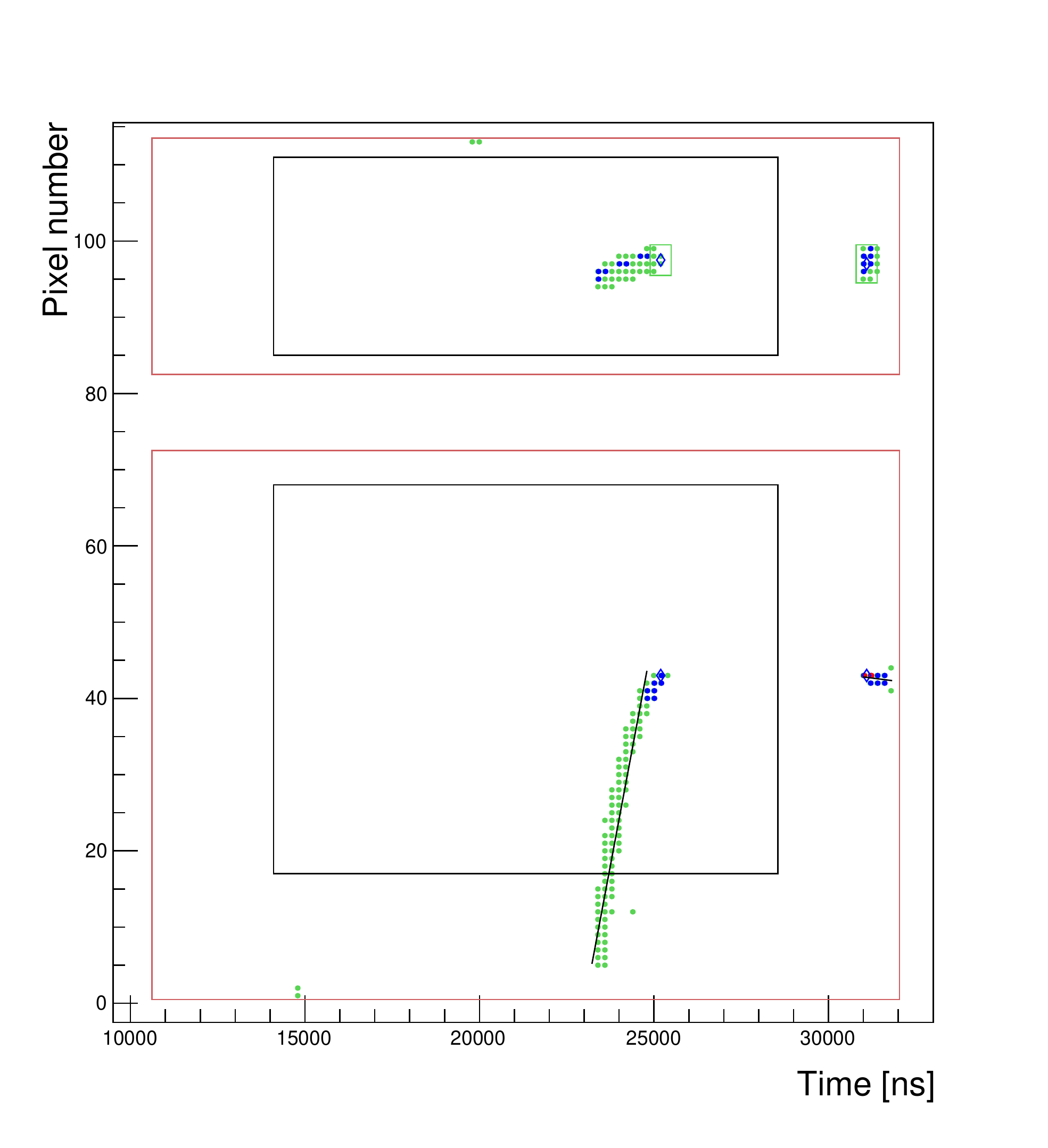}
  \caption{A typical $Z>1$ nuclear muon capture event in the discriminated TPC
  data, as presented in the MuCap event display.  
  The signature of the capture was a time-delayed 
  cluster of EH and/or EVH pixels near the same anode as the muon stop.}
  \label{fig:evdisplay-capture}
  \end{center}
\end{figure*}

\begin{figure*}[tp]
  \begin{center}
  \includegraphics[width=0.9\textwidth]{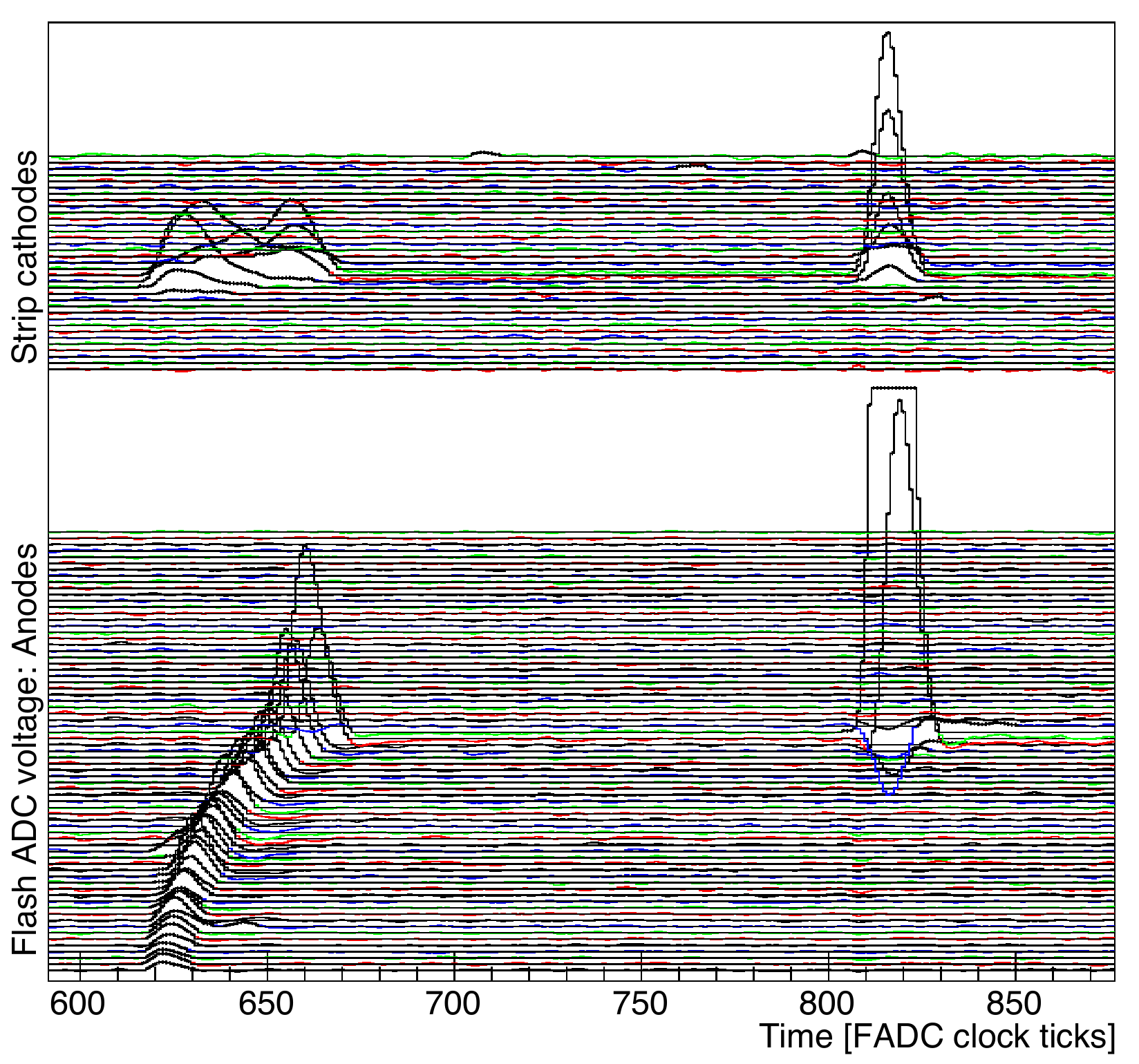}
  \caption{FADC data from the same $Z>1$ capture event shown in
  fig.~\ref{fig:evdisplay-capture}; the anodes and cathodes are arranged 
  in the same manner as in that figure.}
  \label{fig:fadc-display-capture}
  \end{center}
\end{figure*}

	When a muon underwent nuclear capture by a $Z>1$ element inside the TPC, 
	the recoiling nucleus could deposit enough energy to trigger the EH and EVH 
	thresholds, depending upon the nucleus and direction of recoil. The signature 
	of such an event was a time-delayed cluster of EH and/or EVH pixels on or 
	near the anode where the muon stopped.  The discriminated TDC400
and FADC data
from a typical $Z>1$ capture event are shown in figs.~\ref{fig:evdisplay-capture}
and~\ref{fig:fadc-display-capture}, respectively.

To identify $Z>1$ capture events in the TPC data, we searched for EH and EVH pixels 
in the time following each 25-$\mu$s-pileup-protected muon stop that was
unaccompanied by a decay electron candidate within a 0--20~$\mu$s period. The 
capture search was limited to a 1--10~$\mu$s time interval and a $\pm$3 anode 
interval~(inclusive) from the anode corresponding to the muon stop, as illustrated in fig.~\ref{fig:evdisplay-capture}. The 1~$\mu$s 
gap between the end of the muon track and the start of the capture search was necessary 
to avoid events in which the muon stop signal and an EH capture signal overlapped.  
The finite time window was sufficient to observe a maximum of 85\% of captures, assuming 
the majority of contaminants were nitrogen and oxygen.  As a control on the capture 
search, a similar search was performed in the opposite direction---that is, leftwards, in 
the time preceding the muon stop.  The ``impurity capture'' yields 
($N_{\rm captures}/N_{\rm stops}$) from these control searches in the $\mu^-$
data were nearly identical to the ``capture'' yields in the $\mu^+$ data, supporting the 
veracity and reliability of the control results.

The $\mu^-$ impurity capture yields from the TPC data tracked in time with 
external events in the gas system---namely, changes in the operating status of 
CHUPS, and the intentional addition of impurities for calibration purposes.  
When CHUPS was turned on, the TPC capture yield asymptotically 
approached a level as low as 3.4~ppm, corresponding to a minuscule humidity 
concentration of about 9~ppb.  The yield was observed to shift up 
or down with changes 
to the rate at which the gas was purified and circulated.  
Materials inside the pressure vessel continued to outgas, despite the prolonged 
baking of the hydrogen system prior to the run; the primary contaminant was  oxygen in the form of water.

Although CHUPS was effective in suppressing $Z>1$ impurities,
nonnegligible levels of contamination were still observed in the
hydrogen gas~\cite{Andreev:2012fj}.  As a result, it was necessary to perform a zero-extrapolation
correction in the offline analysis for the effects of impurities on the muon 
decay time spectrum.  The TPC-based impurity capture yield was the primary 
observable used to perform this correction.  The impurity concentrations  
always remained well below the limits (roughly 30~ppm for nitrogen and 
14~ppm for oxygen) where their effects on the muon lifetime spectrum 
would start to become nonlinear.

\subsubsection{Information derived from the flash ADCs}
\label{sssec:fadc}

The FADCs played an auxiliary but valuable role in the experiment. They were used to calibrate the
anode energy scales and thresholds, study the gain stability over time, and 
obtain spectral information about the capture recoils which the primary
discriminated signals could not provide. The FADCs were triggered by an 
FPGA with a prescaled minimum-bias trigger and two triggers selecting rare capture events
within the central 48 anode wires of the TPC. 
One trigger simply fired on any EVH signals, while the other trigger 
searched for two consecutive EH triggers on the same anode. 
A pulse finding algorithm defined the time, amplitude and energy
of pulses above the software threshold, and these pulses were then matched to
the corresponding TDC signals.  

\begin{figure}[htbp]
  \begin{center}
  \includegraphics[width=\columnwidth]{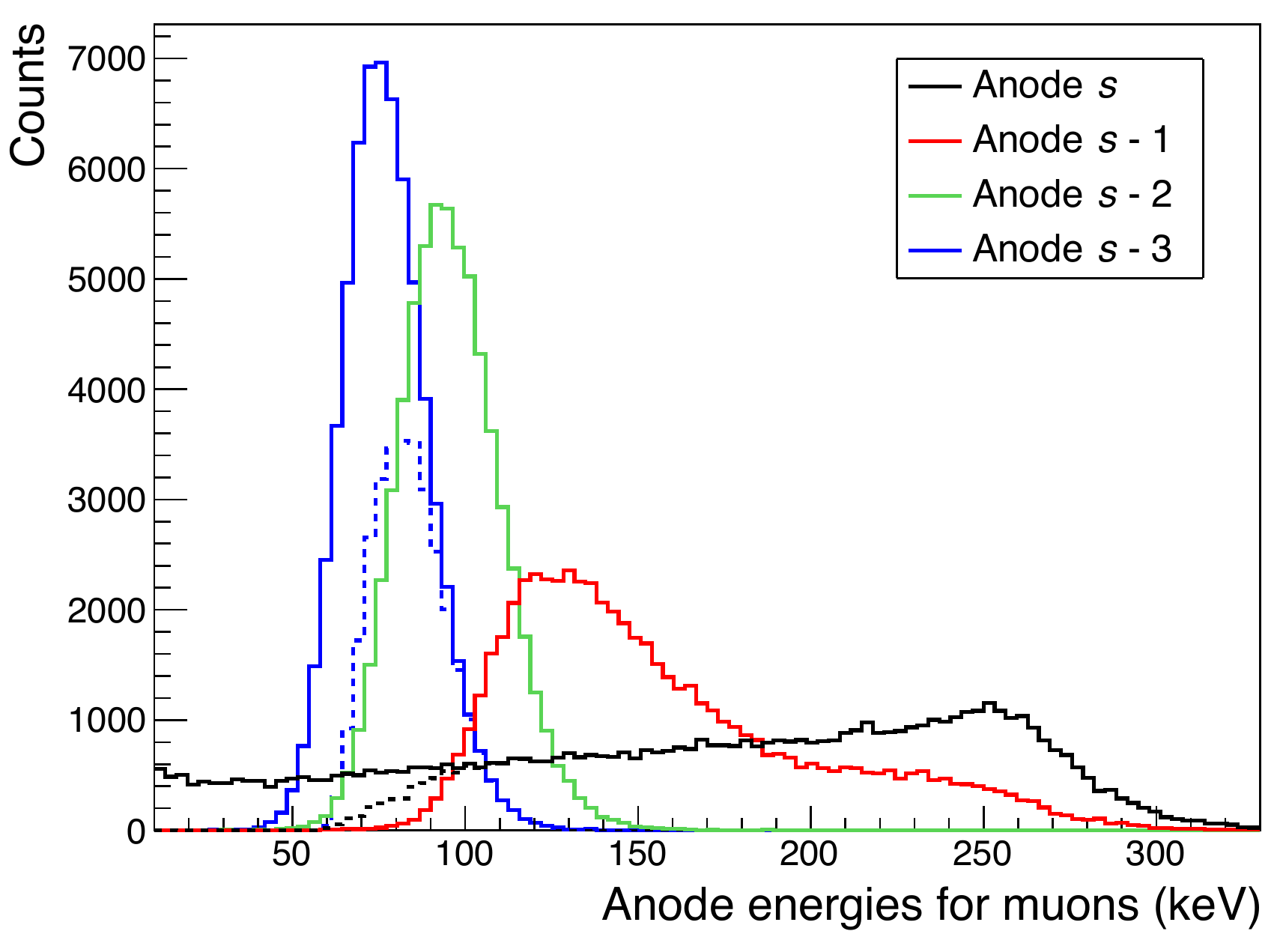}
  \caption{Muon energy deposition in stopping anode $s$, and the first, second and third anodes upstream from it.
  The dashed curves are the energy spectra gated by the EH discriminator for anode $s$ and $s-3$; the EH threshold at $\sim$~75 keV is clearly visible.}
  \label{fig:MuonFADC}
  \end{center}
\end{figure}

The energy spectrum for typical muon stops is 
shown in fig.~\ref{fig:MuonFADC}.  Here anode $s$ denotes the muon stopping 
anode identified by the TDC track finding algorithm, and anodes $s-i$ are 
upstream of that stopping anode by $i$ positions.  The Bragg curve becomes 
flatter with increasing distance to the stopping point,
and as a result the energies deposited on upstream anodes fell within 
relatively narrow ranges, as indicated by the well-defined peaks for anode $s-2$ and $s-3$
in fig.~\ref{fig:MuonFADC}. 
The energy deposited on anode $s$, however, could range from 0 to $\sim$~250 keV, 
depending how deeply the muon penetrated the 4~mm region covered by the anode. 
The well-defined energy depositions in anodes $s-3$ through $s-8$ were used used to calibrate the energy scale for all
 anodes by adjusting their values to predictions from GEANT and SRIM~\cite{Ziegler}.
 The data in fig.~\ref{fig:MuonFADC} was compiled from muon stops on the central anodes.

\begin{figure}[htp]
  \begin{center}
  \includegraphics[width=\columnwidth]{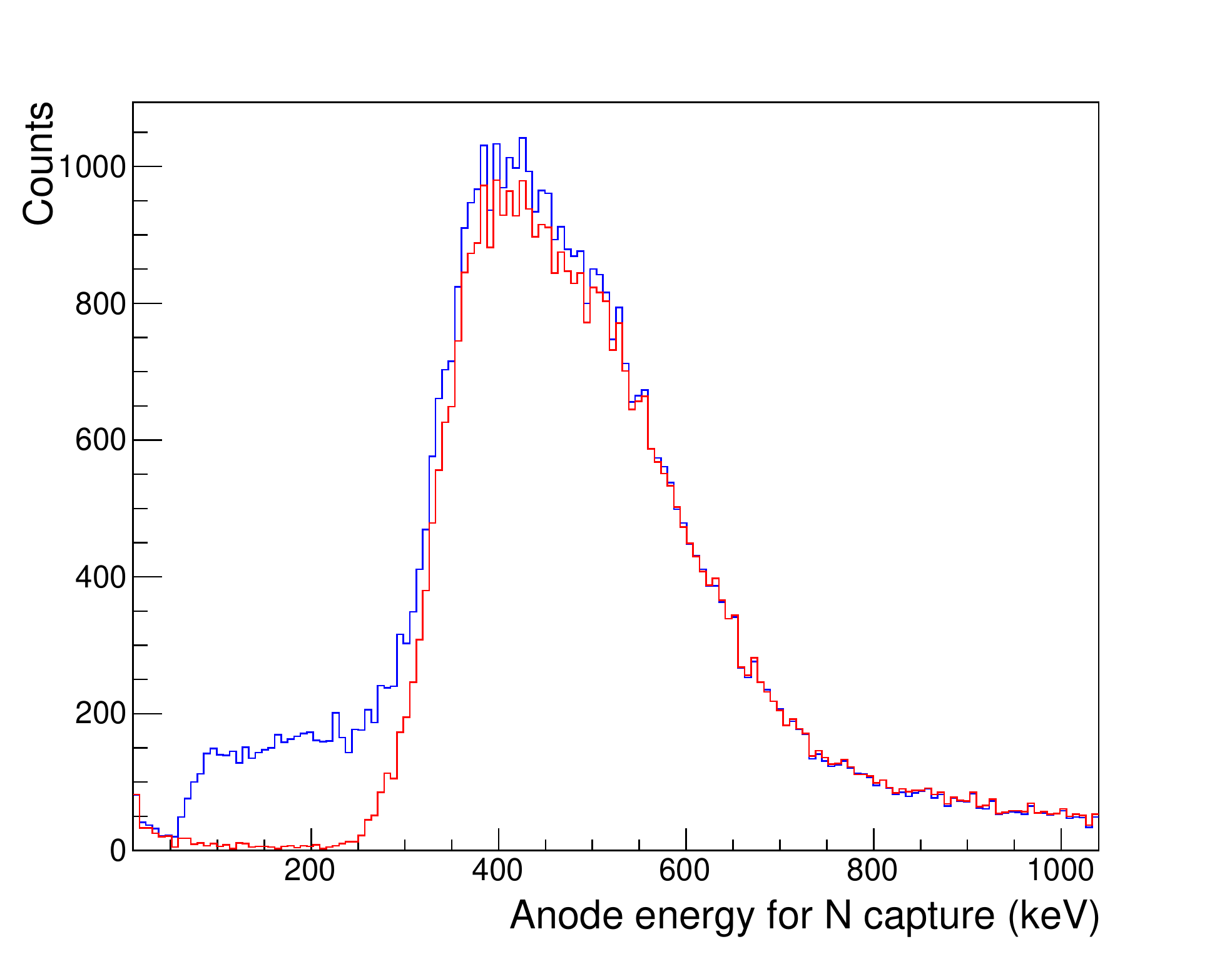}
  \caption{Energy deposition of selected impurity capture events in 
nitrogen-doped data, as determined from FADC signals. The blue curve shows the 
distribution selected by the EH capture trigger, while the red curve shows the same distribution after 
requiring the EVH trigger; the EVH threshold 
around $\sim$~340~keV is clearly visible.}
  \label{fig:CaptureFADC}
  \end{center}
\end{figure}

 The TDC thresholds were determined for individual wires from the half-height of the threshold curve, obtained by dividing the
energy spectra gated by thresholds EL, EH and EVH by the ungated spectra. The value of EL differed depending on the gas gain during each individual data-taking period, as it was set above noise. The EH and EVH thresholds were set according to the muon energy deposition and rarely changed.
Typical values obtained by the FADC-based calibration are given in 
sect.~\ref{ssec:electronics}. The average EH threshold can
also be read off fig.~\ref{fig:MuonFADC}, where the dashed spectra show the threshold effect on anode $s$ and $s-3$, 
respectively.  The location of the EVH threshold can similarly be determined from fig.~\ref{fig:CaptureFADC}, which shows the energy spectra from capture events that fired the EH and EVH capture trigger in a nitrogen-doped run.
The selection was based on a similar capture search algorithm as described
in sect.~\ref{sssec:impurities}. 

For the impurity capture analysis, the stability and reproducibility of the EVH threshold was most important.  The impurity
correction was based on the yield of capture events with energies above the EVH setting; the threshold had to be stable within a data-taking
run so that the yield measured with a doped calibration mixture 
 could be applied to the pure protium  data. 
The nitrogen calibration was measured in the 2004 and 2006 runs but not
in 2007, so threshold consistency across the runs was also 
required.   The EVH setting remained stable to 
10\% during each run and within 10\% between them.
Based on the observed capture energy spectra, 
this was sufficient to apply the impurity correction.

\section{Conclusions}
\label{sec:conclusions}

The MuCap experiment employed a first-of-its-kind, high-pressure hydrogen-gas 
TPC as the central detector in a precision measurement of the rate of nuclear
muon capture in protium.  The device was crucial to the collection of
$1.2 \times 10^{10}$ fully tracked muon decay events in hydrogen.
The resulting measurement of the lifetime of 
the $\mu p$ singlet atom had a precision of 16~ppm, allowing the nuclear muon capture 
rate to be determined at the 1\% level.  The experiment ultimately confirmed 
precise theoretical predictions from chiral perturbation theory for the value of the weak nucleonic pseudoscalar coupling $g_P$~\cite{Andreev:2012fj}.

The MuCap experiment also pioneered the use of the TPC as a chemical impurity 
detector, observing nuclear recoils from capture on elemental impurities in 
the hydrogen.  In this role it enabled precise corrections to be made to the 
$\mu p$ lifetime for effects arising from the presence of nitrogen and oxygen 
contamination in the hydrogen gas at the 10 ppb level.  The TPC provided 
continuous, in situ monitoring of the impurity level, in contrast to 
procedures like gas chromatography that operate on discrete samples.

The expertise gained from the operation of the MuCap TPC has been 
vital to the development of a new ${\rm D}_2$-filled TPC for use in the 
MuSun experiment, which seeks to measure the rate of muon capture in
deuterium.  This next logical step in the progression of muon capture 
experiments requires deuterium gas cooled 
to 30~K~\cite{Andreev:2010wd,doi:10.1146/annurev-nucl-100809-131946}.
The overall program systematically explores the physics and operation of 
ultra-pure hydrogen TPCs, covering a density range 10-70 times higher than 
standard chambers at atmospheric pressure.

\section*{Acknowledgments}
The authors wish to thank all members of the MuCap collaboration for
allowing the experimental data to be used in this publication, as well 
as all staff who made possible the construction of this instrument.
This work was supported by U.S. National Science Foundation, 
the U.S. Department of Energy, CRDF Global, PSI, 
the Russian Academy of Sciences,
and the Grants of the President of the Russian Federation.

\bibliographystyle{epj}
\bibliography{mucap-tpc-nima-paper.bib}

\end{document}